\documentclass[preprint2]{emulateapj}

\shorttitle{carbon nanoparticles} \shortauthors{J\"ager et al.}

\begin{document}
\title{Spectral properties of gas-phase condensed fullerene-like carbon nanoparticles from far-ultraviolet
to infrared wavelengths}

\author{C. J\"ager\altaffilmark{1}}
\affil{Max-Planck-Institut f\"ur Astronomie, K\"onigstuhl 17,
D-69117 Heidelberg and Institut f\"ur Festk\"orperphysik,
Helmholtzweg 3, D-07743 Jena, Germany}
\email{Cornelia.Jaeger@uni-jena.de}
\and
\author{H. Mutschke}
\affil{Astrophysikalisches Institut und Universit\"ats-Sternwarte
(AIU), Schillerg\"{a}sschen 2-3, D-07745 Jena, Germany} \and
\author{Th. Henning}
\affil{Max-Planck-Institut f\"ur Astronomie, K\"onigstuhl 17,
D-69117 Heidelberg, Germany} \and
\author{F. Huisken}
\affil{Max-Planck-Institut f\"ur Astronomie, K\"onigstuhl 17,
D-69117 Heidelberg and Institut f\"ur Festk\"orperphysik,
Helmholtzweg 3, D-07743 Jena, Germany}
\altaffiltext{1}{Corresponding author}

\begin{abstract}
Carbon solids are ubiquitous material in the interstellar space.
However, the formation pathway of carbonaceous matter in astrophysical environments as
well as in terrestrial gas-phase condensation reactions is not yet understood.
Laser ablation of graphite in different quenching gas atmospheres such
as pure He, He/H$_2$, and He/H$_2$O at varying pressures is used to synthesize very
small, fullerene-like carbon nanoparticles. The particles are characterized
by very small diameters between 1 and 4\,nm
and a disturbed onion-like structure. The soot particles extracted
from the condensation zone obviously represent a very early stage of
particle condensation. The spectral properties have been measured from the far-ultraviolet (FUV)
($\lambda$=120\,nm) to the mid-infrared (MIR) ($\lambda$=15\,~$\mu$m).
The seed-like soot particles show strong absorption bands
in the 3.4\,~$\mu$m range. The profile and the intensity pattern of the 3.4\,~$\mu$m
band of the diffuse interstellar medium can be well reproduced by the measured
3.4\,~$\mu$m profile of the condensed particles, however, all the carbon which
is left to form solids is needed to fit the intensity of the interstellar bands.
In contrast to the assumption that onion-like soot particles could be the carrier
of the interstellar ultraviolet (UV) bump, our very small onion-like
carbon nanoparticles do not show distinct UV bands due to ($\pi-\pi$*) transitions.
\end{abstract}


\section{Introduction}
The interstellar extinction law contains a number of, more or less,
strong spectral features which are attributed to absorption bands of
different organic and inorganic dust materials
\citep{Henning:98,He:03}. Unfortunately, the identification of most
of these features remained uncertain. The strongest extinction
feature at a wavelength of 217.5 nm (for reviews see
\citet{HeSchnai:98,Fitzpatrick:03}), which corresponds to an
oscillator strength per H nucleus of $\Delta$f $\approx$
9.3$\times$10$^{-6}$ must be caused by a relatively abundant dust
material composed of the most abundant elements in space, like
carbon, silicon, oxygen, iron, and magnesium. In the diffuse
interstellar medium, the position of the 217~nm absorption is
remarkably invariant for different lines of sight, but the width of
the peak varies by around 25~\%. Carbonaceous material containing
single and double bonds can show electronic transitions in the UV.
Based on calculations of the spectral properties from the optical
constants of bulk graphite, a first assignment of the UV bump
carrier to small nano-sized graphite particles has been suggested by
\citet{Stecher:65}. However, in the laboratory, nano-sized graphite
grains with a very definite graphite structure could not be
observed \citep{Huffman:75,Huffman:88}. Even though there were some concepts to attribute the
feature to small PAHs or mixtures of large PAHs
\citep{Donn:68,Joblin:92,Beegle:97}, solid carbon grains showing
disordered internal structures accompanied by a degree of
hydrogenation are generally accepted as the best candidates for band
carriers \citep{Mennella:96,Duley:98,Schnai:97,Mennella:99}. More
recently, carbon onions have been proposed as band carriers of the
UV bump by different authors
\citep{Wright:88,Wada:99,Tomita:02,Chhowalla:03}. However, most of
the proposed laboratory analogs show deviations of their UV band
positions compared to the interstellar UV bump.

The ''original'' carbon dust is being formed in the envelopes of
carbon-rich late-type stars as nanometer-sized particles
\citep{DoHe:95,Waters:03,Andersen:03,Kwok:04}, but the 217\,~nm band
has never been detected in these objects. However, observations of
these stars in the UV range are rather limited since such stars are
too cool to be observable in this range. Post AGB stars and
planetary nebulae are generally brighter in the UV, however, the
spectral change of the IR features from the post AGB to the
planetary nebulae stage points to a chemical and structural
alteration accompanied by an aromatization process of the
carbonaceous dust \citep{Hony:03}. The hydrogen-rich object HD
89353, that has been identified as a post AGB star, does not show a
UV band \citep{Buss:93}, which could be a hint for the existence of
hydrogenated amorphous carbon material (HACs) \citep{Muci:94}. In
hydrogen-poor objects including post AGB stars \citep{Buss:89},
planetary nebulae \citep{Greenstein:81}, and R Coronae Borealis
variable stars \citep{Drilling:89,Drilling:97}, UV bands between 230
and 250\,~nm were observed, which are typical of rather graphitized
carbonaceous material \citep{Muci:94}. The UV spectrum recorded for
the post AGB star HD 44179 recently presented by \citet{Vijh:05a} do
also not show a 217.5\,~nm absorption peak but, instead, a very
broad hump with a maximum near 200\,~nm (5\,~$\mu$m$^{-1}$). These
observational facts indicate that a chemical or physical processing
of the condensed carbon particles has to occur during the transfer
from the circumstellar shells to the diffuse interstellar medium or
during their stay in this medium.

Generally, carbon nanoparticle composites (soot) consist of primary
particles with average particle sizes between a few and approximately
100\,~nm depending on the mechanism of formation. The structure of the
primary particles is complex and must be explained on an atomic
scale as well as on a mesoscopic scale (short-range and medium-range
order). Carbon atoms can form different types of hybridization, sp,
sp$^2$, sp$^3$, and also mixtures of them which determine the kind
of bondings between the C-atoms and possible heteroatoms such as
oxygen. There are various models \citep{Heidnr:68,Dnetb:93} for the
description of the internal structure of soot particles on a mesoscopic
scale. These models are mainly based on the assumption that the
structure-forming units of carbon nanoparticles consist of very
small crystalline, turbostratic graphitic regions, the ''Basic
Structural Units'' (BSU) \citep{Oberlin:90}. If the sample becomes
more amorphous, the number of stacked layers decreases until only
one layer is left. Simultaneously, the number of condensed rings in
the plane also decreases, which means that the crystallite sizes in
c- and also in a-direction are reduced. A more recent model for the
description of the internal structure of carbon black particles is
the paracrystalline model \citep{Dnetb:93} which predicts strongly
disturbed and bent, concentrically arranged, graphene layers.

During the last few years, many more structures of primary soot
particles have been detected in the laboratory, including nanotubes
and more or less perfectly formed onion-like particles
\citep{He:03}. In these particles, the deviation from the plane
structure of the graphene layers is clearly visible and correlated
with the formation of mixed hybridization states. \citet{Had:93}
investigated the correlation between the s-character of the $\pi$
bonding which describes the deviation from the planar symmetry and
therefore the strength ratio between the $\sigma$ and $\pi$ bonds
and the state of curvature of graphene layers.  The author found
that the hybridization of carbon atoms in bent layers increases to
values between 2.0 and 2.278. The latter is the hybridization value
of C$_{60}$.

The UV absorption of carbon nanoparticles is caused by electronic
transitions between the bonding and antibonding orbitals
\citep{Green:90}. The ($\sigma$--$\sigma$*) transitions are expected
to produce a band in the far UV peaking between 60 and 100 nm,
whereas the ($\pi$--$\pi$*) transitions are providing an absorption
maximum located in the range between 180 and 280~nm. The
incorporation of hydrogen into the internal structure of carbon black
leads to an increase of the fraction of sp$^3$ hybridized carbon.
The position of the ($\pi$--$\pi$*)
transition is extremely sensitive to very small changes of the
preparation conditions, which correspond to small variations in the
internal electronic structure of the carbon nanoparticles
\citep{Rob:87,Mich:98}. There has been only a limited number of
systematic experimental investigations on the relation between the
internal structure and their UV absorption behavior
\citep{Jaeger:99,Llamas:07}. The width of the peak depends also on
the state of agglomeration \citep{Rou:96,Schnai:97}.

A significant property of carbon soot is the hydrogen content which,
on the one hand, influences the internal structure of the grains, and
on the other hand, permits the observability in the infrared (IR) range. The
incorporation of hydrogen into the carbon structure can induce a
reduction of the mean graphene layer length that can be accompanied by an
increase of sp$^3$ hybridization of carbon atoms. Hydrogen bound to
carbon atoms occurs in different functional groups such as in
$\equiv$CH, in aromatic or aliphatic =C--H, or in saturated
aliphatic --CH$_3$, --CH$_2$, or --CH groups. Aromatic =C--H groups
can be observed mainly at 3.3, 8.6, and between 11 and 14.3\,~$\mu$m,
whereas the diagnostically relevant aliphatic --CH$_x$ groups mainly
absorb around 3.4, 6.9, and 7.25\,$\mu$m. There are additional IR bands of these
groups at smaller wavelengths, but they occur in IR ranges where other
groups like --C--C--, --C=C--, C--OH, and --C--O--C-- also absorb and
thus prevent a discrimination and identification of individual peaks.
Macroscopic properties like shape, size, and agglomeration state
influence the long-wavelength tail of the FIR absorption
\citep{Stog:95,Mich:95,Jaeg:98,Quinten:02}.

The IR bands of hydrogen-containing functional groups have a high
diagnostic importance since they trace the processing of the solid-state
organic carbon components in different astronomical
environments. Aromatic and aliphatic IR bands preferentially caused
by varying C--H groups can be observed in the IR spectra of post AGB
stars and protoplanetary nebulae \citep{Kwok:01,Hony:03,Hrivnak:07, Goto:08}
pointing to a mixture of both components in such
objects. In contrast, planetary nebulae only show aromatic IR
bands which can be explained by a complete aromatization of the
material. Further chemical alteration of the carbonaceous material can
be observed in the diffuse interstellar medium (DISM) where only a
strong absorption band at 3.4\,~$\mu$m appears. A detailed analysis
of the DISM in the IR range between 2.5 and 10\,~$\mu$m has been
performed by \citet{Pendleton:02}. From the spectral signatures,
they derived a mean composition of the interstellar carbon material
in the DISM characterized by a hydrocarbon which contains only little
nitrogen or oxygen and which is composed of aromatic and aliphatic forms. The
authors have shown that the 3.4\,~$\mu$m IR band of different
carbonaceous dust analogs indicates a remarkable similarity to the
3.4\,~$\mu$m IR profile of the DISM. The hydrogenated amorphous carbon materials
produced by resistive heating of carbon rods in He/H$_2$ atmospheres
\citep{Schnai:97}, by laser-ablated carbon rods and subsequent hydrogenation
\citep{Mennella:99}, and by plasma-enhanced chemical vapor deposition of methane
\citep{Furton:99} were found to yield the best agreement with the observational
data and to satisfy all spectroscopic criteria for the comparison.
However, the spectral coincidence of the interstellar 3.4\,~$\mu$m bands with
that of a carbonaceous dust analog is not sufficient for the
identification of the interstellar carbon dust component. The
intensity of the IR band is also a crucial factor for an exact
identification. \citet{Furton:99} and \citet{Mennella:02}
demonstrated that a relevant carbonaceous dust analog must have a mass
absorption coefficient $\kappa$ at 3.42\,~$\mu$m of not less than
1400\,~cm$^2$g$^{-1}$ to reproduce the intensity of the interstellar
3.4\,~$\mu$m bands.  For such a material, around 80\,~ppm of the
interstellar carbon relative to hydrogen is required  which is just
the amount of carbon that is left for the formation of solid carbonaceous dust materials
\citep{Snow:95,Snow:96}. Carbonaceous materials with $\kappa$ values
larger than 1400\,~cm$^2$g$^{-1}$ could match the intensity of the
interstellar 3.4\,~$\mu$m band with less than 80\,~ppm carbon.
However, \citet{Furton:99} stressed that carbon materials with
much higher hydrogen content should exhibit blue-green
photoluminescence. Indeed, blue luminescence was discovered in a proto–planetary nebula,
the Red Rectangle (RR), surrounding
the post-AGB star HD44179 and it was attributed to fluorescence of small, neutral PAHs
containing 3--4 rings \citep{Vijh:04,Vijh:05a}. It was assumed
that the small molecules were shielded from the harsh interstellar radiation by a dense circumstellar disk
which enabled their survival. Later, blue luminescence was detected in several
reflection nebulae illuminated by stars having temperatures between 10\,000 and 23\,000\,~K and showing
aromatic IR emission features \citep{Vijh:05b}.

Gas-phase condensation in the laboratory is a process
which is similar to the astrophysical condensation process of particles in late-type stars.
The evaporation of the graphite by laser ablation leads to the formation of small carbon
clusters which can react with hydrogen atoms in the quenching
gas atmosphere to form hydrocarbon molecules and solid hydrocarbons.

Condensation temperatures and pressures in circumstellar environments are rather uncertain.
Generally, temperature-pressure profiles for carbon stars are difficult to obtain
and depend on various model assumptions, for instance, mass loss rates, gas density
outflow velocity, and dust formation. Pulsations of the helium burning shells
affect the atmospheres and can cause shock fronts which result in density/pressure
fluctuations \citep{Nowotny:05}. Recent modelings of AGB star atmospheres give pressures
between 1 and 800\,~dynes/cm$^2$ (1$\times$10$^{-6}$--0.8\,~mbar) for stars with an
effective temperature of 4000\,~K \citep{Lederer:06}.

In condensation calculations, the pressure in the condensation zone is assumed to be between
10 and 3$\times$10$^3$\,~dynes/cm$^2$ (0.01-3\,~mbar) \citep{Lodders:95}.
The authors modeled condensation sequences for C-rich circumstellar envelopes
and demonstrated that, for graphite, the condensation temperature is not very
sensitive to the pressure but to the C/O ratio. At a C/O ratio of 2
(for example valid in IRC+10216) and a pressure larger than 3\,~mbar,
TiC condenses prior to graphite as it was found in meteoritic grains.
At smaller C/O ratios, this condensation sequence is
found at lower pressures (between 0.003 and 0.3\,~mbar) \citep{Lodders:99}.
The coincidence between the calculated and experimentally found condensation
sequences can confirm the assumed condensation conditions and pressure ranges.

In this paper, we present an investigation of the internal structure
of carbon nanopowder, produced by laser ablation and extracted from the
reaction zone in a very early state of condensation,
in relation to its spectral behavior from the FUV up to the IR.
Pulsed laser ablation and
condensation of particles in a quenching gas atmosphere can be considered as
a high-temperature condensation process that can be employed to produce
carbon nanoparticles built of strongly bent graphene layers which is
comparable to a defective onion structure. Defective onion particles
were supposed to have special spectral properties and were discussed
as possible carriers of the interstellar extinction bump \citep{He:03,Tomita:02}.
Therefore, such particles are of great interest as laboratory dust analogs.
In order to study the influence of hydrogen on the structure of the
fullerene-like particles and on the spectral properties of these materials,
we have varied the hydrogen content in the carbon particles.
The understanding of the spectral properties of such grains from the far UV to the IR range
is essential for the further search for carriers of circumstellar and interstellar bands.
In Sect.\,\ref{experi}, the generation of the
samples as well as the techniques for analytical characterization
are described. Section \ref{result} contains the results of the
analytical and spectral characterization and continues with a
discussion of the results.

\section{Experimental\label{experi}}
\subsection{Sample production\label{prod}}
\begin{figure}[htp]
\epsscale{1.00} \plotone{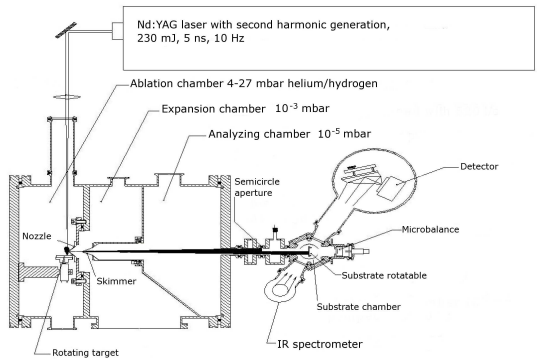} \caption{Molecular beam apparatus
equipped with a laser ablation source, particle extraction,
deposition chamber, and FTIR spectrometer.} \label{laserabl}
\end{figure}

Carbon soot particles were produced by pulsed laser ablation of graphite
and subsequent condensation in a quenching gas atmosphere at
pressures between 3.3 and 26.7\,~mbar. A scheme of the experimental
setup is shown in Fig.\,\ref{laserabl}. Nd:YAG laser pulses with a
wavelength of 532\,~nm (second harmonic) were used to evaporate carbon
from the rotating graphite target (Ringsdorff Spektralkohle). The
duration of the laser pulses was 5\,~ns with pulse energies between 50
and 240\,~mJ. Additionally, defocusing of the laser beam allowed to
vary the power density on the target.
The applied power densities varied between
2$\times$10$^8$--9$\times$10$^9$\,~W\,cm$^{-2}$ resulting in
temperatures of more than 4000\,~K in the condensation zone
\citep{Jaeger:08}. The quenching gas pressures have been changed in
order to influence the chemical and structural properties of the
condensates. The applied low pressure regimes between 3.3 and
6.0\,~mbar are comparable to the pressure conditions for dust
condensation in AGB stars with a C/O ratio of 2, where a pressure
larger than 3\,~mbar is necessary to condense TiC prior to graphite
\citep{Lodders:99}. In our condensation experiments, helium,
hydrogen, water, and mixtures of He/H$_2$ and He/H$_2$O served as
quenching gases. The He/H$_2$ and He/H$_2$O flows were kept constant
during the condensation experiments by employing flow controllers
and amounted to 5:3 and 50:1, respectively.

The condensation of carbon particles is
caused by collisions between the evaporated carbon atoms and
clusters in a supersonic expansion of the hot plasma. Further
cooling is achieved by collisions with rare gas atoms and
the molecules of the admixture. In order to avoid strong structural processing and
agglomeration of the originally condensed grains, the particles were
extracted from the condensation zone through a nozzle and skimmer
to form a freely propagating particle beam. Finally, the carbon grains were
deposited on KBr and CaF$_2$ substrates for IR
and UV/VIS spectroscopy, respectively. The distance between the
condensation zone and the nozzle was of the order of a few
millimeters and was kept constant for all experiments. The
experimental setup allows us to record IR spectra of the deposited
particles without exposure to air. The amount of deposited carbon
particles on the substrate was determined by means of a quartz
microbalance, i.e. a part of the particle beam was directed to a
microbalance in order to measure the thickness of the particle
layer.

\subsection{Analytical characterization of the condensed carbon grains\label{anal}}
The internal structure of carbon primary particles was investigated
by high-resolution transmission electron microscopy (HRTEM). The
term 'primary particles' is used for the condensed individual soot
particles composed of subunits such as bent graphene layers. The
internal structure of these particles is characterized by the types
of bonds (short-range order), the lengths of the subunits, and
distances between them (medium-range order). These analytical
characterizations are very important since the internal structure
determines the spectral properties of the material. HRTEM has been
performed using a JEOL JEM 3010 microscope equipped with a LaB$_6$
cathode operating at an acceleration voltage of 300\,~kV. For this
purpose, the carbon particles were directly collected on TEM grids
from the extracted particle beam in the third chamber (analyzing
chamber). The internal structure of the particles as a function of
the experimental condensation conditions could be quantified
statistically by performing various image analyses which are
specified in a previous paper \citep{Llamas:07}. The HRTEM
micrographs were Fourier-transformed (FT) to reveal any
'periodicity' in the structure, and the intensity profiles of the
computer-generated diffractograms have been applied to measure the
mean distances d$_{002}$ between the graphene layers for the chosen
sample area. The sizes L$_a$ of these graphene layers were
determined within sample areas of 15\,~nm diameter for all evaluated
HRTEM images. These image parts were skeletonized by filtering the
images using ring-shaped masks, which were appropriate to eliminate
the majority of periodicity without physical sense. An inverse FT
from the filtered frequencies was employed to generate the
skeletonized image which was used to measure the sizes of the
graphene layers.

Electron energy loss spectroscopy (EELS) was applied using a Gatan
Imaging Filter, which was attached to a 200 kV HRTEM (Philips CM 200
FEG). The EEL spectra have been used to determine the mean ratio
between sp$^2$ and sp$^3$ hybridized carbon in the particles. The
method is described in detail in previous papers
\citep{Jaeg:98,Jaeger:99}. For quantitative structural analyses, we
have employed the core-loss measurements, where the (1s--$\pi^*$)
transitions at 286~eV and the broad (1s--$\sigma^*$) transitions
above 290~eV were used to distinguish between the sp$^2$ and sp$^3$
hybridization states of carbon. The sp$^2$/sp$^3$ ratio was
quantitatively determined by fitting the core-loss spectra with
Gaussian profiles and computing the integral intensities of the
$\pi$ to $\sigma$ bands. The calculated ratio of $\pi$ to $\sigma$
electrons was compared with the ratio determined in a standard
carbon soot sample. This standard sample was a soot material
produced by resistive heating of graphite rods and subsequent
condensation of nanoparticles in a helium atmosphere. The
sp$^2$/sp$^3$ ratio in this sample was measured by EELS and, in
addition, by $^{13}$C-NMR, and both results were in very good
agreement \citep{Jaeger:99}. Bent graphene layers that can be
observed in the HRTEM images represent mixed hybridization states
which means that the (1s--$\pi^*$) band is shifted to higher
energies \citep{Aja:93}.

In-situ IR transmission spectra of the deposited particles
were recorded using a Fourier Transform IR spectrometer (Bruker 113 v) in the
wavelength range between 2 and 20\,$\mu$m for the KBr or between 2 and
12\,$\mu$m for CaF$_2$ substrates. The fraction of hydrogen in the soot
grains has been determined from the analysis of the IR spectra in the
range of the -C--H stretching vibrational bands at 3.4 $\mu$m
(3000\,--\,2800\,cm$^{-1}$). The relation between the integrated strength
of the saturated aliphatic --C--H bands and the hydrogen content in the
samples is based on a method proposed by \citet{Jac:96}.

FUV/UV/VIS transmission spectroscopy in the wavelength range between
115 and 1000 nm was performed on the soot samples deposited
on CaF$_2$ substrates by means of a VUV spectrometer (LZ Hannover)
covering the range between 115 and 230 nm and a standard grating spectrometer
(Perkin Elmer Lambda 19) operating between 190 and 3000\,~nm.

\section{Results and Discussions}\label{result}
\subsection{Structural characterization}\label{struc}
\begin{figure*} \epsscale{1.00} \plotone{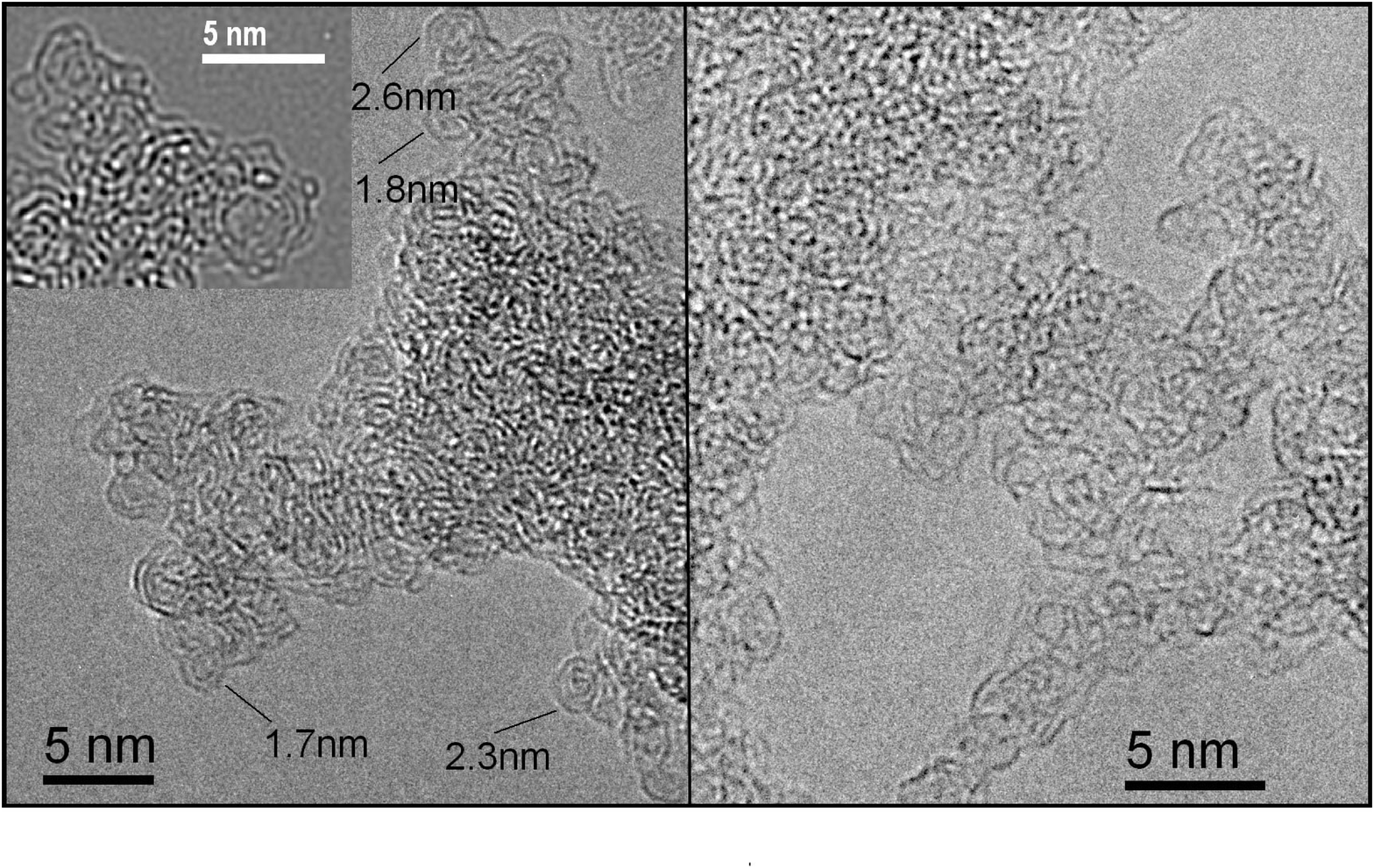} \caption{HRTEM
images of soot particles produced by laser ablation in 26.7\,~mbar
He (left image) and in 3.3\,~mbar He/H$_2$ quenching gas atmospheres
(right image). The inset in the left panel shows characteristic
fullerene structures that can be observed in the samples. The very
small onion-like grains are clearly visible at the periphery of the
three-dimensional aggregates. Some typical sizes of these individual
grains are derived from the micrograph.}\label{HRTEM}
\end{figure*}

The laser ablation/condensation setup combined with the particle beam
extraction of the condensed grains from the hot condensation zone,
allowed to extract the freshly condensed carbon grains in a very
early state of formation avoiding strong processing of the grains in
the hot condensation zone. In order to investigate the size,
shape, and structure of the carbon grains, we have applied HRTEM.
For these purposes, we have directly collected the carbon grains on a
TEM grid from the extracted particle beam. HRTEM images have shown
that the condensed grains are very small, fullerene-like particles (see
Fig.\,\ref{HRTEM}). These particles are composed of small, strongly bent
graphene layers with varying lengths (L$_a$) and distances between
these layers. All observed grain structures can be described as
amorphous showing the typical structureless halo in the Fourier-
transformed bright field images. The level of disorder depends on
the employed condensation conditions. To derive the structural
parameters of the carbon particles quantitatively, we have used the
image analysis described in Sect.\,~\ref{anal}. The results of this analysis
are presented in Table\,\ref{Tab_HRTEM} together with the ratio
between sp$^2$ and sp$^3$ hybridized carbon determined from EELS
(expressed as the fraction of sp$^2$) and the hydrogen content
determined from IR spectroscopy.

\begin{table*}[htp] \caption[]{Quantitative structural parameters of the condensed carbon samples.}
\label{Tab_HRTEM} \vspace*{0.2cm}\scriptsize
\begin{tabular}{ccccccccc}
\hline
Sample & Laser pulse energy & Pressure & Quenching  &H/C  & sp$^2$ C& Mean L$_a$ & Largest L$_a$ & Mean d$_{002}$ \\
       & mJ/pulse           & (mbar)   & gas        &     & \%      &    (nm)    &      (nm)     & value (nm)     \\
\hline
 S1    & 240 uf*            &  26.7    &  He        &0.14 &    56   &  0.53-0.6  &  2.7          &   0.43         \\
 S2    & 240 f*             &  26.7    &  He/H$_2$  &0.16 &    57   &  0.49      &  2.2          &   0.51         \\
 S3    & 240 f              &  3.3     &  He/H$_2$  &0.39 &     -   &  0.45      &  1.7          &   0.54         \\
 S4    & 50 f               &  6.0     &  He/H$_2$  &0.50 &    48   &  0.42      &  1.8          &   0.59         \\
 S5    & 240 uf             &  6.0     &  He/H$_2$  &0.52 &    48   &  0.42      &  1.7          &   0.57         \\

 S6    & 2.4 f              &  4.0     &  He/H$_2$O &0.57 &     -   &  0.43      &  1.8          &   0.57         \\
 S7    & 2.4 f              &  13.3    &  He/H$_2$O &0.41 &     -   &  0.44      &  1.7          &   0.55         \\
 S8    & 0.8 f              &  13.3    &  He/H$_2$O &0.16 &     -   &  0.49      &  2.0          &   0.48         \\
\hline
\end{tabular}
\tablenotetext{*}{The abbreviations uf and f stand for unfocused and
focused laser beam, respectively.}
\end{table*}

The derived mean sizes of the graphene layers L$_a$ are very small
for all carbon samples. However, one can observe a decrease of the
layer size with decreasing pressure and decreasing laser power. The
more hydrogen is incorporated the smaller is the mean L$_a$.

The sp$^2$/sp$^3$ ratio increases with growing quenching gas pressure.
The incorporation of hydrogen in carbon nanoparticles leads to
smaller sizes of the subunits and higher content of sp$^3$ hybridized carbon
along with higher curvature radii of the graphene layers.

There are strong variations in the hydrogen content, but only small
differences in the mean graphene layer sizes and distances between
the graphene layers and in the overall structure of the grains.
The soot samples produced in mixed He/H$_2$ atmospheres have shown
that the insertion of hydrogen is highest for those produced at
low pressure (see S1 and S2 compared to S3-S5 in Tab.\,~\ref{Tab_HRTEM})
and with low laser power or unfocused laser beam (see S4 and S5).

Generally, the applied laser power densities between 1.2 and
9.2$\times$10$^{9}$\,~Wcm$^{-2}$ and the associated temperatures should be high enough to dissociate the
molecular hydrogen efficiently. The vibrational temperature of the
laser-induced plasma, generated by laser evaporation of a graphite target
in a 10\,mbar He atmosphere with power densities between
0.5-2$\times$10$^9$\,~Wcm$^{-2}$ was found to range between 4000 and 6000\,~K
\citep{Iida:94}. At small laser powers, the number of hydrogen atoms becomes low enough
to produce not only volatile organic molecules but also solid
nanograins. The latter assumption is supported by the experimental
fact that, at lower laser power, the soot yields are higher than at
high laser power. Additionally, the level of hydrogen incorporation
into the structure is influenced by a competitive reaction of hydrogen release at high
temperatures resulting in optimum conditions for H insertion at low pressure and
laser power.

For the soot samples condensed in mixed helium/water
atmospheres, we found that the hydrogen content in the
carbon structure is very high, even at very low water partial
pressures. In contrast to the He/H$_2$ samples, the hydrogen content
is increasing with growing laser
power. This is due to the fact that the number of generated hydrogen
atoms from water molecules is much lower compared to the condensation
process in He/H$_2$ atmospheres even if all water molecules are
dissociated in the laser field. Consequently, the limited
number of hydrogen atoms in these atmospheres leads to the formation
of solid grains instead of the formation of organic molecules.
The oxygen is partly inserted as carbonyl
(--C=O) groups in the structure.

\subsection{Spectroscopic characterization}\label{spec}
\begin{figure}[htp]
\epsscale{1.00} \plotone{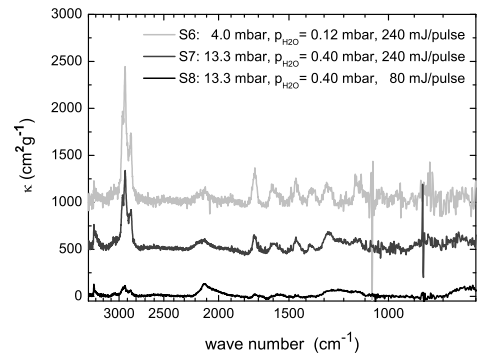} \caption{IR spectra of soot
produced by laser ablation in He/H$_2$O quenching gas atmospheres at
different pressures and laser energies.} \label{IR_H2O}
\end{figure}

The IR mass extinction coefficient spectra of soot condensed in
He/H$_2$O atmospheres, derived from the baseline-corrected IR
transmission spectra, are presented in Fig.\,\ref{IR_H2O}. The mass
extinction coefficient has been calculated from transmission $T$ by
$\kappa=-lnT(d\times\rho)^{-1}$ where $\rho$ and $d$ represent the
density of the carbon grains and the thickness of the layer,
respectively. The density was determined to be 1.55\,~g\,cm$^{-3}$
by a pyknometer measurement whereas the layer thickness has been
measured with the quartz microbalance.

The IR measurements have been performed in situ on the freshly deposited carbon particle films
in a vacuum of 2$\times$10$^{-6}$\,~mbar in order to prevent the
adsorption of organic molecules on the surface of the dust particles from air.
This is a general problem of many of the previously published measurements of
the optical response function of carbon soot particles which are known to be efficient
adsorbers.
The IR band attribution is given in Tab.\,\ref{Tab_IR_bands}.
\begin{table}[htp] \caption[]{Assignment of the IR bands to the functional
groups present in the carbon soot materials.} \label{Tab_IR_bands}
\vspace*{0.2cm} \footnotesize
\begin{tabular}{cc}
\hline
IR band position & Assignment of the bands      \\
 cm$^{-1}$       &                              \\
\hline
3300 - 3310      &   $\equiv$C--H stretching \\
3060 - 3050      &    =C--H stretching \\
3000 - 2800      &   --C--H stretching \\
2110 - 2125      &   --C$\equiv$C-- stretching \\
1720 - 1730      &   --C=O stretching \\
1593 - 1605      &   --C=C-- stretching \\
1455 - 1465      &   --C--H deformation \\
1360 - 1380      &   --C--H deformation \\
1270 - 1280      &   --C--C-- stretching, --C--H deformation\\
1120 - 1145      &   --C--C-- stretching, --C--H deformation\\
850  -  860      &   =C--H out of plane, 1 H, 2 adjac.H\\
735  - 745       &   =C--H out of plane, 3 to 5 adjac.H\\
\hline
\end{tabular}
\end{table}
Only very weak aromatic C-H stretching bands at around
3060\,cm$^{-1}$ have been detected in the samples, but bands between
1605 and 1610\,cm$^{-1}$, observed in all samples, confirm the
existence of aromatic --C=C-- in the carbon structure. The hydrogen
is mainly incorporated as saturated --CH$_2$ or --CH$_3$ groups
probably acting as bridges between bent aromatic subunits.  Besides
the strong --C--H stretching vibrations at 3.4\,$\mu$m, the typical
--C--H deformation bands around 1460 and 1375\,cm$^{-1}$ have been
detected. Additionally, small --C=O features around 1730\,cm$^{-1}$
have been measured. Since we were not able to deposit carbon layers
with thicknesses larger than 170\,nm, we obtained very noisy spectra
in the long-wavelength range of the IR. Therefore, only very noisy
aromatic out-of-plane bending =C--H vibrational bands could be
observed at around 740 and 850\,cm$^{-1}$. The IR spectra show high
mass extinction coefficients in the 3.4\,$\mu$m range up to
1550\,cm$^{2}$g$^{-1}$ for the samples with the highest content of
hydrogen in the structure.

\begin{figure}[htp]
\epsscale{1.00} \plotone{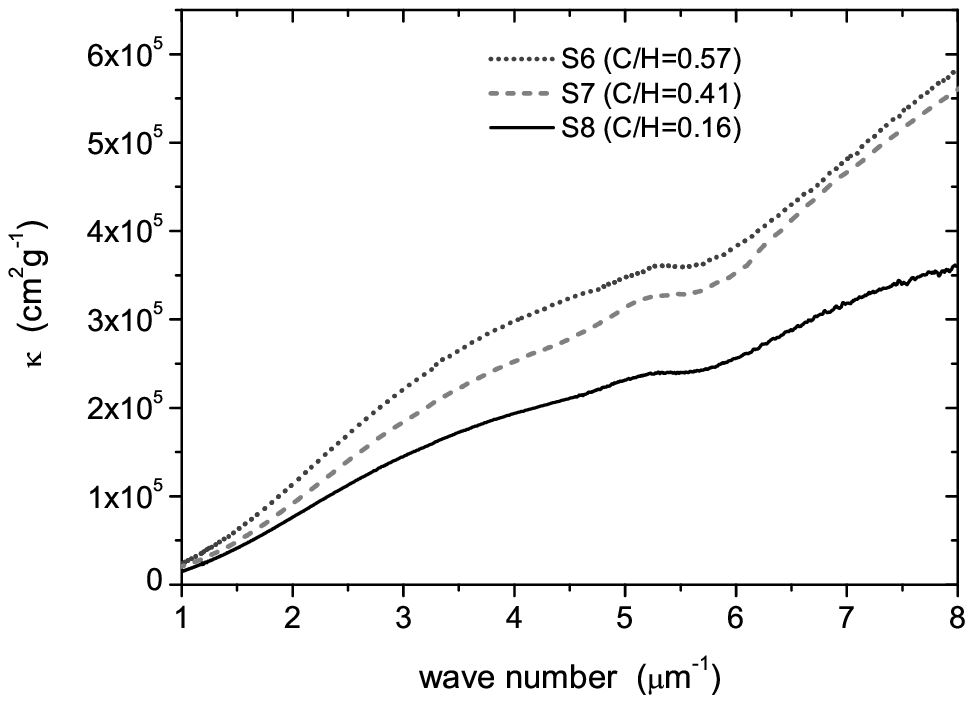} \caption{UV/VIS spectra of the soot
samples produced in He/H$_2$O quenching gas atmospheres at varying
pressures. IR spectra are shown in Fig.\,~\ref{IR_H2O}.}
\label{UV_H2O}
\end{figure}

\begin{figure}[htp]
\epsscale{1.00} \plotone{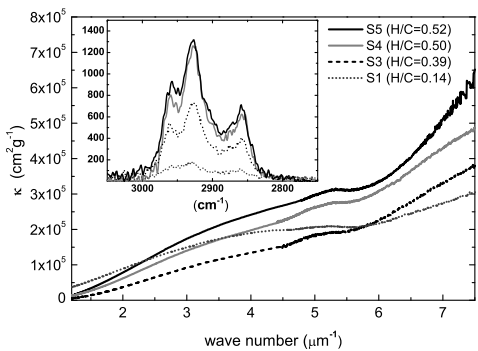} \caption{FUV/UV/VIS spectra of soot
produced by laser ablation in He and He/H$_2$ quenching gas
atmospheres at varying pressures and laser energies. The preparation
conditions of the samples are given in Table\,~\ref{Tab_HRTEM}. The
inset shows the corresponding IR spectra in the 3.4\,~$\mu$m range
of the saturated --CH$_3$ and --CH$_2$ groups.} \label{fuv}
\end{figure}

The corresponding extinction spectra in the FUV to VIS spectral
range of the soot particles produced in He/H$_2$O atmospheres are
depicted in Fig.\,\ref{UV_H2O}. Although we have determined a
considerable amount of sp$^2$ hybridized carbon in the samples, the
UV spectra do not show a distinct bump due to ($\pi-\pi$*)
transitions.
An identical spectral behavior, both in the IR and the FUV/UV/VIS
ranges, has been found for the carbon nanoparticle layers produced
from condensation in He/H$_2$ atmospheres. The UV/VIS spectra of
these condensates shown in Fig.\,~\ref{fuv} do not exhibit distinct
$\pi-\pi*$ bands either. The change in the slope at about
5.8\,$\mu$m$^{-1}$ indicates a very broad ($\pi-\pi$*) band, merging
with the long-wavelength wing of the strong and broad
($\sigma-\sigma$*) band located in the far UV.

\begin{figure}[htp]
\epsscale{1.00} \plotone{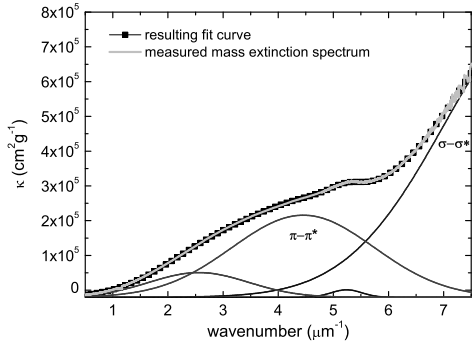} \caption{Deconvolution of the
UV/VIS spectrum of the soot sample produced in 6.0\,~mbar He/H$_2$
atmosphere having the highest hydrogen content of H/C= 0.52 (S5)
using four Gaussian profiles. The resulting fit curve is marked with
filled quads. The weak band at 5.2\,~$\mu$m$^{-1}$ is caused by the
absorption of --C=O groups.} \label{Fit}
\end{figure}

Fig.\,\ref{Fit} shows the deconvolution of one of the UV/VIS spectra into
Gaussian profiles. Similarly to \citet{Llamas:07}, we employed four Gaussians
to obtain an accurate fit of the extinction spectrum. The strong oscillator
around 9\,~$\mu$m$^{-1}$ accounts for the ($\sigma-\sigma$*) electronic
transitions whereas the broad Gaussian profile centered around 4\,~$\mu$m$^{-1}$
describes the ($\pi-\pi$*) transitions.
The weak Gaussian at approximately 2.5\,~$\mu$m$^{-1}$ is interpreted as
a plasmon peak due to some larger but strongly bent graphene layers.
The width of the ($\pi-\pi$*) main band and the appearance of the plasmon band
accounts for a strong disorder in the carbon structures due to a broad
distribution of curvatures and lengths of graphene layers in the small
fullerene-like carbon grains. For the sample with the highest hydrogen content, the ($\pi-\pi$*)
and the ($\sigma-\sigma$*) main bands were found to be located at
4.45 and 8.79\,~$\mu$m$^{-1}$, respectively. With decreasing hydrogen content
in the sample, the ($\pi-\pi$*) band is slightly shifted to
smaller wave numbers (3.84\,~$\mu$m$^{-1}$) whereas the ($\sigma-\sigma$*)
transition band remains at the same position (8.8\,~$\mu$m$^{-1}$).
The measured mass extinction coefficients were found to be dominated by absorption.
Scattering measurements have been
performed at soot samples produced by laser pyrolysis consisting of very similar
small fullerene-like particles \citep{Llamas:07}. For these samples,
an approximately linearly increasing scattering contribution ranging between
2\% for wave numbers at 2\,~$\mu$m$^{-1}$ and 11\% for 4\,~$\mu$m$^{-1}$
was measured.

At this point, we have to address a possible influence of
nanoparticle distribution and clumping on the CaF$_2$ substrates on
the UV spectral properties. Despite we have combined our gas-phase
condensation with a particle beam extraction, we could not
completely avoid the agglomeration of our originally condensing very
small fullerene-like particles. In a condensation process which is
caused by a high supersaturation of vapor, one has to expect the
condensation of a large number of very small particles (nuclei).
Further particle growth is exclusively due to particle coagulation.
This is exactly the process we can observe in the HRTEM images of
the particles, which were always directly deposited on the TEM
grids from the extracted particle beam in the third (analyzing) chamber.
The individual particles are very small (less than 3-4 nm).
The largest particle agglomerates are in a range of 15 nm. The
morphology of the particle agglomerates on the substrate can be
understood as a porous layer of rather fractal agglomerates. The
formation of very dense and large grain agglomerates can be excluded
since we have measured only a very low scattering.

Particle clustering and interaction of particles with the substrate can
increase the width of the UV bump. Unfortunately, we cannot exactly
calculate the influence of particle agglomeration on the broadening
of the UV band since we do not have the optical constants of the
material. However, agglomeration effects have been calculated and
measured for carbonaceous particles by several authors
\citep{Schnai:97,Quinten:02}. Strong effects appear only in very
elongated particles, such as prolate spheroids. The formation of
clusters should not completely prevent the formation of a distinct
UV band, which was also shown in previous UV measurements of soot
produced in gas-phase condensations from resistive heating
of graphite rods in quenching gas atmospheres without particle beam
extraction.

The weak band at around 5.2\,~$\mu$m$^{-1}$ (190\,nm) does not change
in position with the structure of the
carbon materials. An origin of this band can be the absorption of
--C=O groups which could be identified in small amounts in
the IR spectra of nearly all carbon soot samples. Electronic transitions
of carbonyl groups have been found to be located between electronic ($\pi-\pi$*) and
($\sigma-\sigma$*) transitions \citep{Braun:05}.
Carbonyl groups bound to carbon can be removed by gentle
annealing. Therefore, we have heated a few samples to temperatures
of around 600\,~K for 1-2 hours or to 523\,K for 8-10 hours. As a
result of the gentle heating of the sample, the small bands at around 190\,~nm
became much weaker, supporting the band assignment to --C=O groups.

\begin{figure}
\epsscale{1.00} \plotone{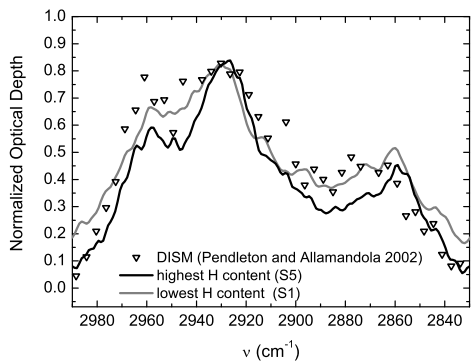} \caption{Comparison of the
3.4\,~$\mu$m profile of two soot materials with the 3.4\,~$\mu$m
profile of the diffuse interstellar medium.} \label{ISM}
\end{figure}

The IR mass extinction coefficients in the spectral range of
the --CH$_x$ stretching vibrations at 3.4\,~$\mu$m of
the soot samples, condensed in He/H$_2$ quenching gas
atmospheres, are shown in the inset of Fig.\,\ref{fuv}.
The sample with the highest hydrogen fraction shows
a mass extinction coefficient of about 1400\,cm$^{2}$g$^{-1}$.
The 3.4\,~$\mu$m absorption profile compares well
with the 3.4\,~$\mu$m extinction profile of the diffuse
interstellar medium \citep{Pendleton:02} which is shown in Fig.\,\ref{ISM}.
However, using the measured oscillator strengths of the --C--H
vibrational bands, we calculated that, even for the samples with
high hydrogen content and a mass extinction coefficient of
approximately 1400 cm$^2$g$^{-1}$, nearly 80 ppm carbon relative to
hydrogen would be consumed to fit the intensity of the interstellar
3.4 $\mu$m profile \citep{Furton:99,Mennella:02}. This is generally
seen as the amount of carbon available for the formation of solid
carbonaceous material \citep{Snow:95,Cardelli:96}.

In the UV absorption model by \citet{Rob:87}, the band structure is determined by
the size distribution of the condensed aromatic islands in the carbonaceous structure
and can be expressed by analyzing the gap energy.
The gap energy $E_g$ can be derived from the optical spectra by means
of the Tauc relation $\sqrt{\epsilon_2 E} = B(E-E_g)$ \citep{Tau:66}
which describes the energy dependence of the imaginary part of the dielectric function
$\epsilon_2=2nk$ above $\kappa \rho$ = 10$^{4}$\,cm$^{-1}$. Instead of $\epsilon_2$,
we have used the absorption coefficient $\alpha$ to derive the gap energy. This is a very
frequently applied method to calculate the gap energy. However, we should stress
that the absorption coefficient is only proportional for a constant
refractive index $n$ which is probably not completely correct for
carbonaceous material.  We have found $E_g$
values between 1.03 and 1.23\,eV for the samples with the lowest and
the highest H content, respectively.
According to the relation $E_g=2 \beta M^{-1/2}$ derived by \citet{Rob:87},
where $\beta$ represents a quantum chemically estimated overlapping energy
between two adjacent p$_z$ orbitals that has an empirical value of 2.9\,~eV, the optical gap energy
of soot is related to the number $M$ of condensed rings in the graphene layers.
Employing this relation, the $E_g$ values between 1.03 and 1.23\,eV correspond
to 32 to 22 condensed aromatic rings in a graphene layer, or to diameters
of the graphene layers between 2.8 and 1.7\,nm (with the diameter of a
single ring being $\approx$0.24\,nm).

The calculated lengths of the graphene layers $L_a$ do not
agree with the mean $L_a$ values determined by HRTEM. Actually, they agree
approximately with the maximum $L_a$ values (see Table \ref{Tab_HRTEM})
representative of only a few strongly bent larger graphene layers seen as parts of
the outer shells of a number of soot particles (cf. Sect. \ref{struc}).
The Gaussian contribution at the smallest wave numbers in the UV/VIS spectrum
(see Fig.\,~\ref{Fit}), probably caused by plasmon excitation in these few
larger graphene layers, can also contribute to the gap energy to a certain extent.

We should note that the discrepancies between the calculated and
measured lengths of the graphene layers can either originate from
the fact that the calculated $E_g$ values are inaccurate, or the
applied model $E_g=2 \beta M^{-1/2}$ is only valid for plane
aromatic units, but not for bent layers and closed fullerene-like
structures. More recent models and actual measurements of gap
energies of large PAHs provide much higher values for E$_g$ than
those calculated by the Tauc relation \citep{Tyu:04,Yang:07}. The
energy gap of PAHs depends not only on the size of the molecules,
but also on their structure, such as the nature of the periphery
\citep{Watson:01}. In any case, this investigation demonstrates
that, in case of a soot material with a wide variation of graphene
layer sizes and curvatures, the optical gap energy may be of limited
value as a structural parameter since it may not reflect the
properties of the dominating structural units.

\subsection{Comparison with other carbonaceous dust analogs}
Our carbonaceous materials produced by using pulsed laser ablation
at high power densities and high condensation temperatures are
structurally the same as the soot produced in our laboratory by
pulsed CO$_2$ laser-driven pyrolysis of hydrocarbons
\citep{Llamas:07}. Therefore, the UV spectral properties are well
comparable. Differences in the IR spectra result from smaller H/C
ratios in the these condensates and a somewhat higher content of
--C=O groups.

Comparisons with condensation products, produced in our laboratory by resistive heating of
carbon electrodes and subsequent condensation of grains in quenching gas atmospheres,
show differences in structure and spectral behavior.
In these condensation experiments, the carbon grains were either clearly larger and contained
longer and less bent graphene layers, or they were characterized by a defective onion-like
structure, but larger grains. The
soot materials showed distinct UV bumps in a range between 196 and 270\,~nm due
to ($\pi-\pi$*) electronic transitions \citep{Schnai:97,Jaeger:99}. The position of the
UV band was found to be sensitive to small changes in the internal structure which is correlated
with modifications of the ratio of sp$^3$ to sp$^2$ hybridized carbon and the content of hydrogen in the sample.
The IR spectra reveal the appearance of analogous bands since they were typical for
functional groups such as --CH$_x$, --C=C--, and --C=O, either present in our samples,
but the intensities of the --CH$_x$ vibrational bands are lower compared to the samples produced
by laser ablation. This indicates a lower hydrogen content in the previous condensates.

Since the pulsed laser ablation and condensation of soot particles is characterized as
a high-temperature (HT) process \citep{Jaeger:08}, other HT condensation processes should provide
comparable materials. The hydrogenated amorphous carbon (HAC) film produced by \citet{Furton:99}
employing a plasma-enhanced chemical vapor deposition reveal very similar structural
parameters as found in our soot samples. The authors did not use electron
microscopy to obtain information on the internal structure of the soot, but they derived the H/C ratio, density,
and hybridization ratio to characterize the sample. The IR mass extinction
coefficient at 3.4\,~$\mu$m is similar to the $\kappa$ value measured for
the sample S5 (H/C = 0.52) in this study, but the derived gap energy is higher than those calculated for our sample.
In the UV/VIS, the material shows similar characteristics with a continuously rising absorption
coefficient in the near-UV.

Laser ablation of graphite employing a pulsed XeCl excimer laser and
condensation of hydrogenated amorphous carbon in different quenching gas mixtures have been performed by
\citet{Scott:96} and \citet{Grishko:02}. The authors analyzed the spectral properties of the condensates
in a broad wavelength range. The UV absorption spectra of the HAC film did not show a UV bump
after deposition at room temperature. Annealing up to 500-600\,~K provoked the appearance of
a band at 4.6\,$\mu$m which is discussed to result from a dehydrogenation and development of aromatic areas of
the size of coronene \citep{Duley:04}. In the IR, the absorption spectra of HACs deposited
at low temperature match well the --C--H stretching bands of the ISM at 3.4\,~$\mu$m \citep{Duley:94}.
Unfortunately, quantitative information on the intensity of these bands is missing.

\citet{Mennella:02} produced particulate carbonaceous dust analogs by pulsed laser ablation of
carbon rods and condensation of the grains in a 10\,~mbar argon atmosphere which can also be
considered as HT condensation process. The grains were exposed to a flux of atomic hydrogen, and
--C--H stretching bands typical for sp$^3$ hybridized carbon atoms have been activated.
The resulting IR spectrum compares well with our IR measurements. Unfortunately, a study of changes
in the internal structure of the condensate and UV/VIS measurements have not been performed at this sample.
Another hydrogen-containing carbon grain material was produced by arc discharge between
two carbon rods in a 10\,~mbar hydrogen atmosphere \citep{Colangeli:95}.
The original grains displayed a featureless UV spectrum at wavelengths longer than 190\,~nm,
as was found for our samples. The corresponding IR spectra are characterized by the typical
--C--H$_x$ vibrational bands at 3.3, 3.4, 6.4, 6.9, and 7.3\,~$\mu$m and a --C=O band at 5.8\,~$\mu$m.
However, the intensity of the 3.4\,~$\mu$m band is quite low and comparable to the one of our
soot material S1 with the lowest hydrogen content. Unfortunately, a characterization of the
internal structure of these grains is missing. Therefore, a classification of the spectral
properties in relation to the inner structure is impossible.

The condensed soot produced by \citet{Herlin:98} employing a continuous-wave (cw) laser-driven
pyrolysis of hydrocarbons, that is considered as a low-temperature (LT) condensation
process, shows a different structure. The authors performed a very careful characterization of the
internal structure of these grains. In contrast to our condensate, they found larger grains
that contain longer, less bent, and more ordered graphene layers. Consequently, the IR bands
are typical of an aromatic
soot condensate with a high ratio between the aromatic =C--H stretching band at 3.3\,$\mu$m
and the aliphatic --C--H stretching bands around 3.4\,$\mu$m. Unfortunately, the UV spectral
properties of the condensate have not been measured. The condensation conditions are very similar
to the condensed soot material produced by cw laser pyrolysis of hydrocarbons described
in \citet{Jaeger:06b} which were found to consist of mixtures of soluble PAHs and solid soot grains.

We would like to emphasize the analogy of
the structure-forming units in our soot particles, that can be considered as fullerene fragments,
to the structural units present in so-called defective onions, even though our soot grains could be
characterized as more disordered.

Carbon onion particles have been discussed to represent a possible carrier
for the interstellar UV bump \citep{He:03}.
\citet{Wada:06} could show that such defective onion-like carbon grains were also found in the
granular and the dark quenched carbonaceous composite (QCC) component condensed by
an eject of a hydrocarbon plasma. The size of the onions ranges from
5 to 15 nm containing an empty core of 2--3\,~nm in diameter.
The onion-like particles are composed of 3 to 15 concentric
shells, that means, they are much larger than our fullerene-like particles.
In contrast to our results, these authors found a distinct
but broad band at around 220\,~nm which is shifted to longer wavelengths upon annealing.
The annealing of the sample was accompanied by an increase of the particle sizes and a
flattening of the carbon layers.

\cite{Tomita:02} produced so-called defective carbon onions
by annealing of nanodiamonds and found an absorption band at 3.9\,~$\mu$m$^{-1}$ for onions
dispersed in water. A theoretical model was developed by the authors to complement the experimental
results. Defective onions of 5\,nm with hollow cores of 0.7\,nm in diameter were found
to fit the interstellar UV bump very well.

Similar theoretical results were published by \citet{Iglesias:04} who could
find a good match between the interstellar 217\,~nm bump and a mixture
of fullerenes and buckyonions with sizes between C$_{60}$ and C$_{3890}$ corresponding to radii
ranging from 0.35 to 2.8\,~nm

Resulting from our experimental study and the different experimental and theoretical approaches
mentioned above, we believe that, for onion particles, the size of
the particles is a crucial parameter and will determine the position and shape of the UV bump.
Large particle sizes shift the UV bump to longer wavelengths
due to flatter graphene layers in the outer part of the particles whereas small particles
can provide a UV band at smaller wavelengths that is comparable to the interstellar UV bump.
Therefore, only onions with a relative monodisperse size can cause a sharp
UV absorption band.
A size distribution of particles would always lead to a broadening
and a shift of the UV absorption due to a distribution of sizes and curvatures of
graphene layers.

The comparison of differently produced gas-phase-condensed carbonaceous matter
makes clear that the spectral properties depend on the production process
and, consequently, on the internal structure of the sample including such parameter as
the mean length of graphene layers, level of bending of the graphene layers, hydrogen content, and
hybridization state of the carbon. In general, the understanding of the absorption behavior
of carbon onions or carbon nanoparticles with more or less
ordered, strongly bent graphene layers in broad wavelength range is essential for the further
search for carriers of the interstellar bands.

\section{Conclusions}
Soot material condensed at high temperatures is characterized
by very small particles due to the high supersaturation of carbon vapor in the condensation
zone resulting in a high number of nucleation seeds and a further particle growth
by coagulation, exclusively. Its structure can be described as fullerene-like soot containing elongated or
symmetric cages which can be interleaved. However, most of the observed structural units
are cage fragments which can be bound to each other either by aliphatic
bridges or by van der Waals forces. The observed graphene layers are strongly bent
inside the small seed-like soot grains leading to a much lower electron density
in the aromatic double bonds. The observed fullerene fragments are very similar
to the structural units observed in so-called defective onions.

We have characterized the soot materials
by several analytical methods such as EELS, FUV, UV, VIS, and IR spectroscopy
in order to determine the ratio of sp$^2$/sp$^3$ hybridized carbon, the content
of hydrogen, and the spectral properties of the generated soot. Structural
parameters (mean lengths of the graphene layers and distances between them) have been derived
directly from the HRTEM images. The condensed soot materials show an upper limit of the
H/C ratios of 0.57 and corresponding peak mass extinction coefficients in the
3.4\,~$\mu$m bands of around 1550\,~cm$^2$g$^{-1}$. The FUV/UV/VIS spectra of soot
with low and high hydrogen content do not show distinct peaks due to electronic
($\pi$-$\pi$*) transitions. The ($\pi$-$\pi$*) transition band is very broad and
rather hidden in the long-wavelength tail of the ($\sigma-\sigma$*) transitions.

There is good agreement between the observed interstellar 3.4\,~$\mu$m profile
\citep{Pendleton:02} and the 3.4\,~$\mu$m absorption of the dust condensate produced in our laboratory.
In addition, our laboratory products fulfill nearly completely the four spectral criteria outlined
by the authors to compare the laboratory product with the interstellar dust material, such as the 3.4\,~$\mu$m
aliphatic --CH$_x$ stretching band profile and subpeak positions, no --OH stretching band near 3.1\,~$\mu$m,
a ratio of the optical depth of the aliphatic --CH$_x$ stretching bands to the optical depth of the carbonyl
band at 5.9\,~$\mu$m of more than 2, and a ratio of the optical depth of the aliphatic --CH$_x$ stretch
features to the optical depth of the --CH$_x$ deformation modes near 6.8 and
7.25\,~$\mu$m of about 5. Nearly 80 ppm carbon relative to hydrogen is necessary to match the intensity
of the interstellar 3.4 $\mu$m
profile for the sample with high hydrogen content and a mass extinction coefficient of 1400 cm$^2$g$^{-1}$
\citep{Furton:99,Mennella:02}.

The bending of the graphene layers in soot particles
causes a change in the electronic structure of carbon, and this
effect was already demonstrated by EELS measurements in the low loss
and the core loss region of the spectra \citep{Jaeger:99}. Here, the electronic
(1s-$\pi$*) transition is shifted to higher energies and the typical
$\pi$ bond shows a growing s-character. Therefore, the position of
the electronic transitions, especially the transition between the
bonding and antibonding $\pi$ orbitals, is determined by the size of
the graphene subunits and/or the degree of bending of these structures.
In our soot particles, the ($\pi-\pi$*) bands are shifted from
4.45\,~$\mu$m$^{-1}$ for the sample with the highest hydrogen content
to 3.84\,~$\mu$m$^{-1}$ in the sample with the lowest content of hydrogen.
The lowering of the hydrogen content in the very small fullerene-like
particles is, at nearly the same degree of bending of the graphene layers
inside the grains, accompanied by a reduction of the mean lengths $L_a$
of these layers. However, a clear relation between the UV band and the degree of bending
and/or the length of the graphene layers, as determined by \citet{Rob:87}
for the plane graphitic subunits, has not been found for soot materials
containing strongly bent graphene sheets. This is due to the difficulties
to determine the level of bending in carbon grains with such
a large variety of different internal structures.

Obviously, the relevant grain formation process is based on the generation of fullerenes and fullerene
fragments in the condensation zone. The precursors for these fragments are probably
chain-like molecules containing C$\equiv$C and C$\equiv$H triple bonds. Hints for the intermediate formation of such
triple bonds can be found in the IR spectra of the soot.
The formation of these cage fragments from the chains
has been modeled by the help of quantum-chemical molecular
dynamics simulations.  High carbon densities are necessary to produce these structures
from C$_2$ \citep{Zheng:05}. At temperatures above 2000\,~K, the molecules quickly combine
to long and branched carbon chains. From the chains, small cyclic structures can develop
with long carbon chains attached. Consequently, fullerene fragments of bowl shape with side chains are
formed \citep{Irle:03}. Such a soot formation process can be of high relevance in supernovae
or in the hot circumstellar environments of carbon-rich stars, such as Wolf-Rayet stars \citep{Cherchneff:00}.

\acknowledgements
This work was supported by a cooperation between the
Max-Planck-Institut f\"ur Astronomie and the FSU Jena as well as by the Deutsche
For\-schungs\-ge\-mein\-schaft (Hu 474/21-1, Mu 1164/4-3). We would like to thank
G. Born for her help in the laboratory.

\end{document}